\documentclass[prl,twocolumn,floatfix,showpacs,superscriptaddress]{revtex4}
\usepackage{amsfonts}
\usepackage{amsmath}
\usepackage{amssymb}
\usepackage{graphicx}

\begin{document}

\title{$\alpha-\gamma$ transition in cerium: magnetic form factor and dynamic magnetic susceptibility in dynamical mean-field theory}
\author{B. Chakrabarti}
\affiliation{Department of Physics \& Astronomy, Rutgers University, Piscataway, NJ 08854-8019, USA}
\author{M. E. Pezzoli}
\affiliation{Department of Physics and Astronomy, Stony Brook University, Stony Brook, New York 11794, USA}
\affiliation{Department of Physics \& Astronomy, Rutgers University, Piscataway, NJ 08854-8019, USA}
\author{G. Sordi}
\affiliation{SEPnet and Hubbard Theory Consortium, Department of Physics, Royal Holloway, University of London, Egham, Surrey TW20 0EX, United Kingdom}
\affiliation{Theory Group, Institut Laue Langevin, 6 rue Jules Horowitz, 38042 Grenoble Cedex, France}
\author{K. Haule}
\affiliation{Department of Physics \& Astronomy, Rutgers University, Piscataway, NJ 08854-8019, USA}
\author{G. Kotliar}
\affiliation{Department of Physics \& Astronomy, Rutgers University, Piscataway, NJ 08854-8019, USA}

\date{\today}

\begin{abstract}
The nature of the elemental cerium phases, undergoing an isostructural
volume collapse transition, cannot be
understood using conventional solid state concepts. Using the
combination of density functional theory and dynamical mean-field
theory, we study the magnetic properties of both the $\alpha$ and the $\gamma$ phases.
We compute the magnetic form factor, and show that it is very close to
free ion behavior in both the local moment $\gamma$ phase as well
as the more itinerant $\alpha$ phase, in agreement with neutron
scattering experiments.  In sharp contrast, the dynamic local magnetic
susceptibility of the two phases is strikingly different.  In the $\gamma$
phase, the sharp low energy peak due to local moment formation and
consequently low Kondo temperature dominates the spectra. In the $\alpha$
phase two broad peaks can be identified, the first is due to Kondo
screening, and the second is due to Hund's coupling.  This shows that
hybridization plays a central role in the $\alpha-\gamma$ transition
in cerium, and that from the point of view of magnetic properties, the
$4f$ electrons are strongly correlated in both phases.
\end{abstract}
\maketitle

The physical mechanism driving the $\alpha$-$\gamma$ phase transition
has puzzled physicists for many years~\cite{ceriumREV}.  Similar to
other elements, the temperature versus pressure phase diagram of
cerium shows multiple structural transitions, where the symmetry of
the structure changes across the phase transition. The
$\alpha$-$\gamma$ transition is exceptional because it is
isostructural, i.e. the atoms retain their ordering in an $fcc$
structure while the volume collapses by 15$\%$ from $\gamma$ to
$\alpha$ phase upon increasing pressure.
Moreover, the $\alpha-\gamma$ transition is accompanied by a dramatic
change in the magnetic susceptibility: the $\alpha$ phase shows
Pauli-like susceptibility, while the $\gamma$ phase has Curie-like
susceptibility. Therefore, most of the theoretical work has focused on
the hypothesis that electronic effects are responsible for the
transition.

Numerous theoretical models were proposed to explain the isostructural
transition in Ce. For example, in the promotional
model~\cite{promotional}, the $4f$ electrons are localized in the
$\gamma$ phase, and are promoted to the $spd$ conduction band in the
$\alpha$ phase.
However photoemission spectroscopy shows little change in the number
of the conduction electrons at the transition.
Johansson proposed~\cite{johansson} that the $\alpha-\gamma$
transition is an example of a Mott transition.  Here the $4f$
electrons undergo a Mott transition, from a non-bonding localized
state in the $\gamma$ phase to a narrow $4f$ band in the $\alpha$
phase, which participates in bonding.  The $spd$ electrons remain
bystanders during the transition.  Upon increasing pressure, a $4f$
localization-delocalization transition occurs with a subsequent loss
of moment and decrease of volume.  In this model, the $4f$ electron
number remains almost unchanged with pressure, so this feature is
consistent with photoemission results.  A different scenario was
proposed by Lavagna et al.~\cite{lavagna} and Allen~\cite{allenCe},
dubbed the Kondo volume collapse theory. Here the $spd$ electrons are
not bystanders as in the Mott scenario. Instead the transition is
connected to the change in the effective hybridization (and thus the
Kondo scale) of the $spd$ electrons with the $4f$ electrons and so
there is a decrease in volume due to the increase of the Kondo
temperature $T_K$.

Dynamical mean-field theory (DMFT)~\cite{rmp} is a modern tool to
understand the physics of strong electron correlations.
The combination of density functional theory (DFT) and DMFT
~\cite{kotliarRMP,hauleLDADMFT} (DFT+DMFT) allows one to
consider  structural effects, electronic effects and the physics of strong
correlations from first-principles. It brings the physics of
f-electron delocalization and f-spd hybridization into a unified
framework.
As a result of many studies over several years, different
aspects of the $\alpha-\gamma$ transition have been considered,
including changes in the density of states~\cite{zolf2001, held2001,
  mcmahan2003}, in the optical conductivity~\cite{hauleCe}, and in the
thermodynamic properties~\cite{amadon,lanata2013}. 

From the theoretical point of view, the magnetic properties of the
volume collapse transition have not been adequately addressed so far.
The static local and the uniform magnetic susceptibility was computed
in Ref.~\onlinecite{streltsov2012} using charge non-self-consistent
DFT+DMFT, but the local susceptibility turned out to be similar in
both phases and the magnetic susceptibility was underestimated due to
negligence of spin orbit coupling, which was shown to be crucial for
proper description of $\alpha$-$\gamma$ transition in
cerium~\cite{lanata2013}. As shown in this letter, the orbital moment
dominates the magnetic moment in cerium, and is quenched when
spin-orbit coupling is neglected.
Recently, compelling neutron scattering experiments~\cite{murani} were
performed, which call for theoretical analysis.
Here we shall revisit the $\alpha-\gamma$ transition from the point of
view of magnetic properties by computing the magnetic form factor
$F_{M}(q)$, the local dynamic susceptibility $\chi(\omega)$ and the
magnetic spectral response $S(q,\omega)$. We shall show that the
magnetic form factor shows free ion behavior in both phases,
indicating that from the point of view of magnetic properties, the
$4f$ electrons are strongly correlated both in the $\alpha$ and
$\gamma$ phases.  The dynamic magnetic susceptibility of the two
phases is very different. It shows a sharp low energy peak at the
characteristic energy, which scales with the coherence temperature of each
phase. Since the coherence scale is directly connected with the
strength of hybridization, this suggests that the hybridization plays
a central role in the $\alpha-\gamma$ transition in cerium.

{\it Model and method.}-- In this Letter, we have performed DMFT+DFT calculations in a charge self-consistent
implementation~\cite{hauleLDADMFT}. For the Kohn-Sham potential, we used the GGA functional
as implemented in Wien2k package~\cite{wien2k}, and continuous-time
quantum Monte Carlo (CTQMC) method to solve the auxiliary impurity
problem~\cite{hauleCTQMC}.  We used Hubbard repulsion $U=6.0$eV,
Hund's coupling $J=0.7$eV, and temperature $T=116K$.  The lattice
constants of the $fcc$ unit cell are $a \approx 4.82\AA$ and $a\approx
5.16\AA$ for the $\alpha$ and $\gamma$ phases respectively.

%
\begin{figure}[!t]
\centering{
\includegraphics[width=0.99\linewidth,clip=]{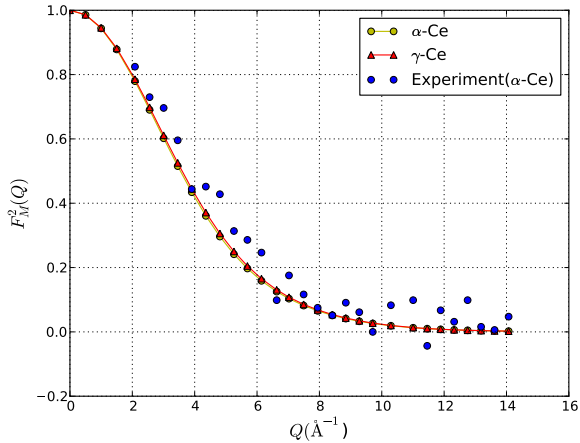}
}
\caption{Momentum transfer dependence of the normalised magnetic form factor F$^2_M$(Q) of $\alpha$-Ce and $\gamma$-Ce. Blue circles show experimental data taken from Ref.~\onlinecite{murani}. 
}
\label{fig1}
\end{figure}

{\it Magnetic form factor.}-- The magnetic form factor $F_M(q)$ is the
Fourier transform of the spatial distribution of the electronic
magnetic moment, here mostly contributed by $4f$ electrons. Thus it is
an ideal observable for determining the nature of the $4f$ electrons.
In particular, the magnetic form factor can determine whether $4f$
electrons are localised or itinerant, as suggested in
Ref.~\onlinecite{hjelm1994}. The idea is the following: band formation
results in the quenching of the $4f$ magnetic moment, especially the
orbital component relative to the spin component. Thus, if the volume
collapse is due to a localisation-to-delocalisation transition, there
should be a dramatic change between the shape of the magnetic form
factor between the $\alpha$ and $\gamma$ phases.  For $\gamma$ cerium,
the measured magnetic form factor has free ion behavior, which is in
good agreement with the computed ionic Ce$^{3+}$ magnetic form
factor~\cite{stassis}.  On the other hand, for the $\alpha$ cerium,
electronic structure calculations predict metal-like behavior for the
magnetic form factor~\cite{hjelm1994}.  However, these calculations
are in striking contrast with recent high-energy neutron inelastic
measurements by Murani et al.~\cite{murani}, which show free ion
behaviour for the magnetic form factor of $\alpha$ cerium as well.

\begin{table}[!t]
\begin{tabular}{c|c|c}
 & $\alpha$-Ce & $\gamma$-Ce \\
 \hline
$\mu_s$ & $-2.3079\times10^{-3}$ & -0.03468 \\
$\mu_L$ & $9.3668\times10^{-3}$ & 0.13841\\
$C_2$ & 1.327 & 1.334 \\
\hline
\end{tabular}
\caption{
Values of the orbital ($\mu_L$) and spin $\mu_S$ magnetic moment as
obtained in our DFT+DMFT calculations under a magnetic field of 10T. The coefficient
$C_2=\mu_L/(\mu_L + \mu_S)$ determines the shape of the form factor in
the dipole approximation and has similar value in both phases. 
}
\label{table1}
\end{table}

Following the formalism described in Ref.~\onlinecite{mariettaMFF}, we
compute the magnetic form factor within the DFT+DMFT framework.
Figure~\ref{fig1} shows the magnetic form factor squared $F_M^{2}(q)$
in presence of an external magnetic field for both $\alpha$ and
$\gamma$ cerium.  The curves are close to each other and display free
ion behavior, typical of a correlated state. This is a consequence of
the electron-electron Coulomb repulsion, and cannot be captured solely
by electronic band structure effects. Our results are in good
agreement with the measured magnetic form factor of $\alpha$ cerium of
Ref.~\cite{murani}, and show that electronic structure calculations,
when combined with the dynamical mean-field theory, have the
predictive power to capture the magnetic response of the $4f$
electrons, and therefore to reconcile theory and neutron scattering
experiment.

To gain a deeper understanding of these results, we resort to dipole
approximation, $F_M = \mu (\langle j_0 \rangle + C_2
\langle j_2 \rangle)$, where $\langle j_k \rangle = \int dr\; u(r)
J_k(qr)$ is the spatial average of the spherical Bessel function
$J_k(qr)$ over the atomic cerium wave function $u(r)$,
$\mu=\mu_S+\mu_L$ is the total (spin plus orbital) magnetic moment,
and $C_2 = \mu_L/(\mu_S + \mu_L)$.  As shown in Table~\ref{table1},
$\mu_L$ and $\mu_S$ have opposite sign because of third atomic Hund's
rule (that is because of spin-orbit coupling and $n_f<1$), and
$\mu_L>\mu_S$, thus $C_2>0$.  The coefficient $C_2$ determines the
shape of $F_M(q)$ and remains basically unchanged across the
$\alpha-\gamma$ transition. It is close to the one expected for a free
Ce$^{3+}$ ion, implying that there is a localised $4f$ electronic
density for both $\alpha$ and $\gamma$ cerium.

\begin{figure}[!t]
\centering{
\includegraphics[width=0.99\linewidth,clip=]{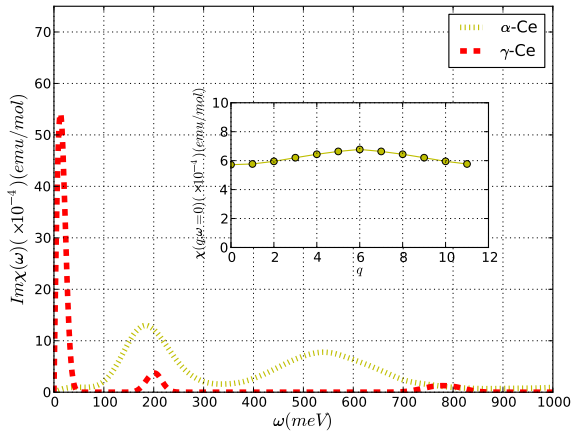}
}
\caption{
  Imaginary part of the local dynamic magnetic susceptibility,
  Im$\chi(\omega)$, for $\alpha$ and $\gamma$ cerium (yellow dotted and red dashed lines  respectively). The inset shows the static
  susceptibility $\chi(q, \omega=0)$ of $\alpha$ cerium as a function of $q$ in the first Brillouin zone. Note that
  the $q$ goes from the points (0,0,0) to (1,1,-1) in 12 uniform
  steps.
}
\label{Susc_both}
\end{figure}
{\it Local dynamic magnetic susceptibility.}-- The magnetic form factor
indicates that both $\alpha$ and $\gamma$ phase are strongly
correlated phases, which is compatible with both Mott and Kondo volume
collapse scenarios, but eliminates the promotional model.  In the Mott
scenario, the two phases are correlated because they lie on slightly
opposite sides of the delocalisation-localisation transition, while in
the Kondo volume collapse picture, the two phases are correlated
because the $4f$ electron moment remains stable across the transition.
However, the Kondo
volume collapse scenario differs from the Mott scenario because it
does not consider the $spd$ electrons to be mere bystanders, but
emphasizes their role in the screening of the local magnetic moments,
via the Kondo effect.  Therefore, the dynamic magnetic susceptibility,
which measures the spatial and temporal distribution of the magnetic
fluctuations, can indicate whether the hybridization plays a key role
at the transition.

In our calculations, we used the CTQMC impurity solver to obtain the
local dynamic  susceptibility $\chi(i\omega_n)$ of $\alpha$ and
$\gamma$ cerium as a function of Matsubara frequencies.  We then
analytically continued the data using maximum entropy method to obtain
Im $\chi (\omega)$ along the real frequency axis.  In Figure
\ref{Susc_both} we show Im $\chi(\omega)$ for both phases.  At small
frequencies, Im $\chi (\omega)$ for $\gamma$ cerium shows a narrow and
intense magnetic peak centered at approximately 10 meV. This feature
has to be expected from the local moment character of electrons in
$\gamma$ cerium.  The position of this peak gives a measure of the
Kondo temperature $T_K$.  For $\alpha$ cerium, this peak shifts to
higher frequency, around 180 meV.  Thus, in going from
$\gamma$ to $\alpha$ phase, there is a shift of magnetic intensity
from low to high energy, signalling a change (precisely, an increase)
in $T_K$.  This is one of the central results of our work.  We
emphasize that, at large frequencies, the overall intensity of Im
$\chi (\omega)$ in the $\alpha$ phase is larger than in the $\gamma$
phase, reflecting the increased hybridization of electrons in the
former phase.

In order to ascertain the nature of the different peaks in the dynamic
susceptibility of $\alpha$ cerium, we performed simulations of the
$\alpha$ phase with different values of spin-orbit coupling (not
shown).  Upon increasing spin-orbit coupling, the peak at low frequency
($\approx 180\,$meV) moves towards $\omega=0$.  This is a feature of
the fact that by increasing the spin-orbit coupling, the effective
Kondo temperature of the system is reduced.  Hence this peak is a
feature of the Kondo coherence energy of the system.  This trend has to
be expected because of the hybridisation between the conduction
electrons and the f electrons. By increasing the spin orbit coupling,
the energy splitting between the $5/2$ and $7/2$ states increases,
therefore fluctuations are hampered and $7/2$ states are less
occupied. It follows that the hybridisation with conduction electrons
decreases as well.  The importance of the spin-orbit coupling has also
been emphasized in the cerium compounds CeIn$_{3-x}$Sn$_x$ and
CePd$_3$~\cite{muraniSO, muraniCePd3, cox1987}.

The second peak ($\approx 600\,$meV) however does not show sensitivity
to the spin-orbit coupling.  To further ascertain the origin of the
second peak, we performed simulations with altered values of Hunds
Coupling $J$ (not shown).  The second peak is always roughly centred
at $\omega=J$ indicating that it represents an excitation of the the
f-electrons in the (non-zero) doubly occupied sector of f-electron
occupancy.

Notice that to crosscheck the validity of our analytic continuation
procedure, we benchmarked our results against a well-known sum rule
for Im$\chi(\omega)$. It is known that
$\dfrac{1}{\pi}\int_{-\infty}^{\infty} n(\omega)Im\chi(\omega) d\omega
= \langle \mu_{z}^{2} \rangle$ where $n(\omega)$ is the Bose
distribution function and $\mu_{z}$ is the magnetic moment along the z
axis, which can be independently extracted in our simulation without
need of analytic continuation.  A good quantitative agreement is
obtained in both phases.

In addition, we have verified that the local dynamical susceptibility
$\chi(\omega)$ is a good representative of the behavior of the
susceptibility $\chi(q,\omega)$ within the first Brillouin zone.  This
is important, because it verifies the so-called "single-ion form
factor dependence" often used to analize experimental
data~\cite{murani}, where the dynamical structure $S(q,\omega)$ is
factorized into momentum dependent form factor $F_M(q)^2$, and
energy dependent structure factor
$S(\omega)=\frac{1}{2}\frac{1}{1-e^{-\beta\hbar\omega}}$Im$\chi(\omega)$,
i.e, $S(q,\omega) = F_M(q)^2 S(\omega)$.  To verify the quality of
this approximation, we have computed the static susceptibility
$\chi(q, \omega=0)$ of $\alpha$ cerium within the first Brillouin zone
using a two particle vertex method, developed in
Ref.~\onlinecite{Park_prl11}.  We calculate $\chi(q)$ using the
Bethe-Salpeter equation $\chi(q)= (\chi_{0}^{-1}(q)-\Gamma)^{-1}$,
where $\Gamma$ is the two particle irreducible vertex, which we sample
within DMFT, and $\chi_0$ is the RPA susceptibility, which we compute
using the full k-dependent LDA+DMFT green's function. Note that within
DMFT, the two particle irreducible vertex $\Gamma$ is local. The inset
of Figure~\ref{Susc_both} shows $\chi(q, \omega=0)$. We can see that
there is no significant variation in the static susceptibility within
the first Brillouin zone, which validates the "single-ion form factor
formula".
\begin{figure}[t]
\centering{
\includegraphics[width=1.2\linewidth,height=90 mm]{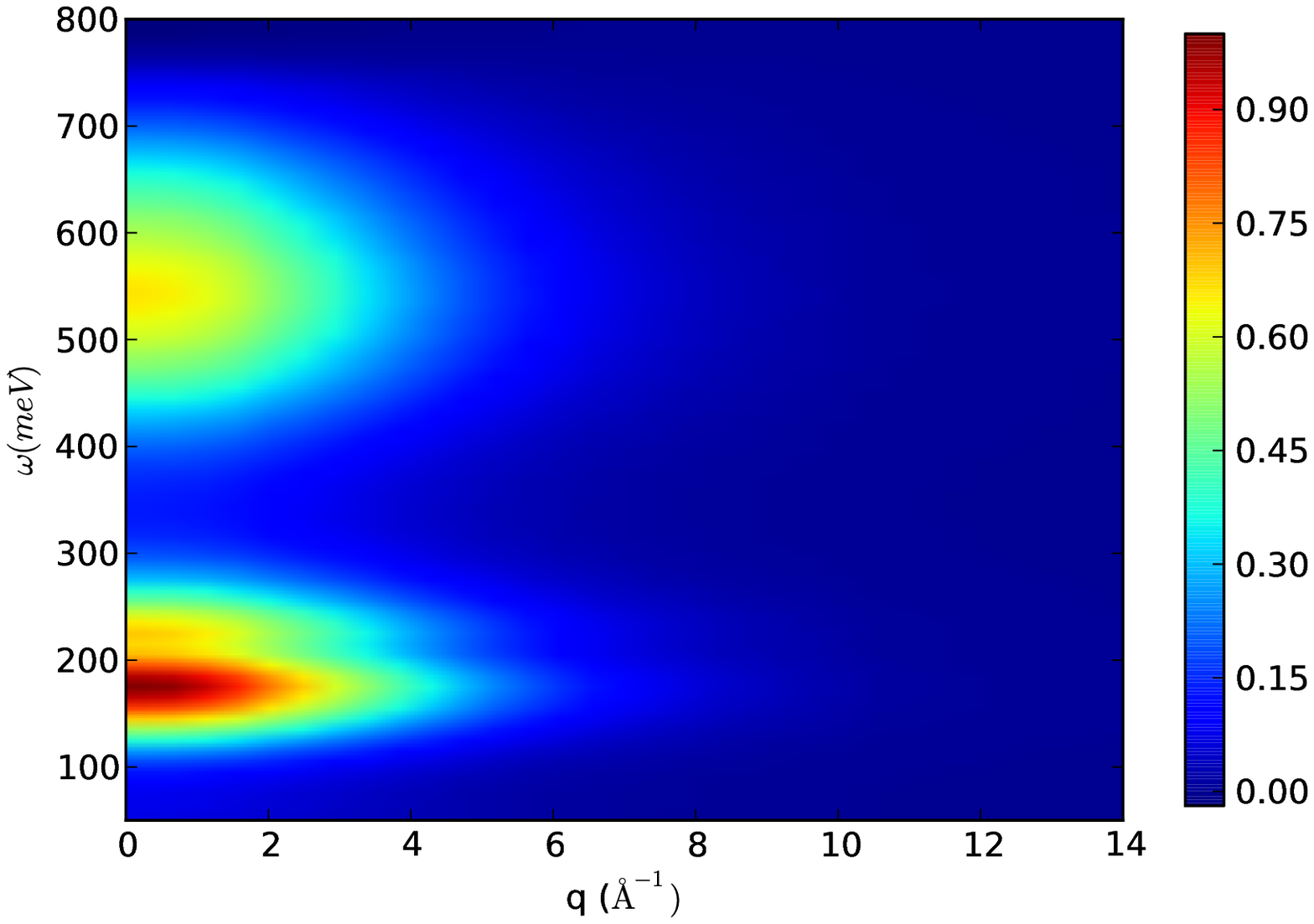}
\includegraphics[width=0.9\linewidth,height=65 mm]{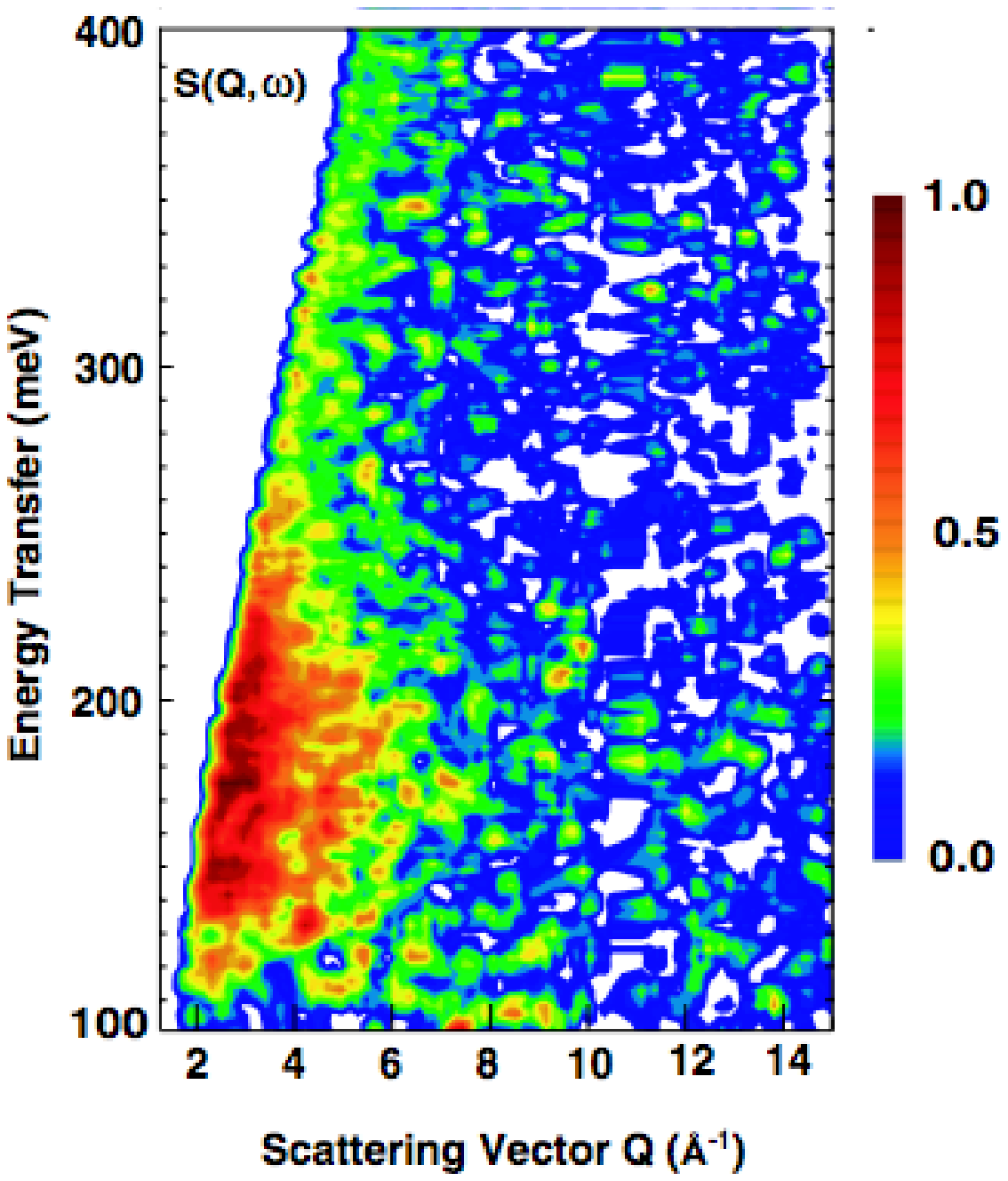}
}
\caption{Difference of $S(q,\omega)$ between $\alpha$ and $\gamma$ phase :
  Top panel shows DFT+DMFT  results. Lower panel shows high energy neutron
  inelastic measurements, taken from Ref~\cite{murani}.
  }
\label{Sq}
\end{figure}

{\it Magnetic spectral response.}-- We then use this formula to
compute frequency dependent $S(q,\omega)$ of both phases.
Figure~\ref{Sq} shows the difference between the $\alpha$ and $\gamma$
magnetic spectral response ($S_\alpha(q,\omega)-S_\gamma(q,\omega)$).
In the lower panel we show the
experimental spectrum~\cite{murani}. There is a good agreement between
the two, particularly in the position of the broad peak assigned to
the Kondo screening in the $\alpha$-phase.  Note that the spectrum
displayed in Fig.~\ref{Sq} becomes negative in the low energy region
where $\gamma$-Ce susceptibility has sharp peak due to local moment character
(see Fig.~\ref{Susc_both}).  This region has been left out of
theoretical (as well as experimental) plot so as to enable better
visualization of the other features.

In summary, we showed theoretically that the neutron magnetic form
factor $F_M(q)$ has a free ion behaviour in both phases, indicating
that the $4f$ electrons remain strongly correlated across the
$\alpha$-$\gamma$ transition.  On the other hand, the local dynamical
susceptibility $\chi(\omega)$ and the magnetic spectrum $S(q,\omega)$
show dramatic changes across the transition, with an energy shift from
lower to higher frequencies, a direct consequence of the increase of
the Kondo temperature $T_K$ in the $\alpha$ cerium.  Therefore, our
data shows that the physics of the volume collapse $\alpha$-$\gamma$
transition in cerium is controlled by the hybridization between the
localised $4f$ and the $spd$ electrons and also establishes the
importance of using different probes and observables to understand
different aspects of the volume collapse transition in cerium.

{\it Acknowledgment.}-- M.E.P. and G.S. thank A. Murani for
enlightening discussions. This work was partially supported by NSF
DMR-0746395 (K.H.) and and DE-FG02-99ER45761(G.K)

\end{document}